\documentclass[showpacs,amsmath,amssymb,amsfonts,twocolumn,prl,floatfix,superscriptaddress]{revtex4} 
\usepackage{graphicx}
\usepackage{bm}
\usepackage[german,english]{babel}

\begin{document}

\title{Construction of analytical many body wave functions for correlated bosons in a harmonic trap}

\author{Ioannis Brouzos}
\email{ibrouzos@physnet.uni-hamburg.de}
\affiliation{Zentrum f\"ur Optische Quantentechnologien, Luruper Chaussee 149, 22761 Hamburg, Germany}

\author{Peter Schmelcher}
\email{pschmelc@physnet.uni-hamburg.de}
\affiliation{Zentrum f\"ur Optische Quantentechnologien, Luruper Chaussee 149, 22761 Hamburg, Germany}

\begin{abstract}
We develop an analytical many-body wave function to accurately describe the crossover of a one-dimensional bosonic system from weak to strong interactions in a harmonic trap. The explicit wave function, which is based on the exact two-body states, consists of symmetric multiple products of the corresponding parabolic cylinder functions, and respects the analytically known limits of zero and infinite repulsion for arbitrary number of particles. For intermediate interaction strengths we demonstrate, that the energies, as well as the reduced densities of first and second order, are in excellent agreement with large scale numerical calculations.
\end{abstract}

\maketitle

\paragraph{Introduction}

Ultracold dilute quantum gases represent an amazingly rich platform for the realization of strongly interacting many-body systems \cite{pethick_smith}. Extensive control of the trapping geometry via external fields as well as tuning of the interaction properties of ultracold atomic ensembles is nowadays routinely possible \cite{bloch,chin}. Theoretical models of quantum many-body systems that were considered as rough idealizations in the past, can now be prepared in a very pure way in order to study, for example, quantum phase transitions \cite{greiner}. One very fundamental model of this type is explored in the present work: an ensemble of bosons at zero temperature, confined to a one-dimensional (1D) harmonic trap and interacting via a repulsive contact potential
of arbitrary strength. Although conceptually simple, this system is in general not analytically solvable. However, the experimental preparation of the trap is possible and the interaction strength can be tuned to any desired value by adjusting the strength of magnetic fields in the vicinity of a Feshbach resonance \cite{chin}, or by employing confinement induced resonances \cite{olshanii}, a tool specific to quasi-1D waveguide-like systems. 

In view of the rich experimental possibilities, many of the ground-breaking theoretical works performed in the 1960's, such as the celebrated Gross-Pitaevskii equation \cite{Pitaevskii}, have attracted large attention. Focusing on 1D systems, and interaction strengths beyond the mean-field regime, two seminal works of this period are of special interest: (i) the Lieb-Liniger solution \cite{lieb} via the Bethe Ansatz for contact interacting bosons in absence of external potential with periodic boundary conditions  (extendable to hard-wall boundaries \cite{hao}) and (ii) the Tonks-Girardeau gas \cite{girardeau} for hard-core bosons mapped to non-interacting fermions (Bose-Fermi map) through the effective Pauli exclusion principle imposed by the infinite repulsion (fermionization). The Tonks-Girardeau gas \cite{kinoshita} and its counterpart for the attractive excited state, the Super-Tonks gas \cite{haller}, were both experimentally realized within the last decade. The Bose-Fermi map works for arbitrary potential geometry, while the Lieb-Liniger solution applies to arbitrary interaction strength;  in the homogeneous case, the two coincide for infinite repulsion. 

In this letter we construct an analytical many-body wave function which describes contact interacting bosons in a parabolic external potential for arbitrary interaction strength. It reproduces the two limiting cases of zero interaction and infinite repulsion for the harmonic trap. Our approach is based on products of functions inspired by the exact solution of the underlying two-body problem. It shows an impressive agreement with results from extensive numerical studies and offers the possibility of extensions to, e.g., higher dimensions or other trap geometries. 

\paragraph{Hamiltonian}

 For the investigation of a 1D trapping geometry one should take into account the experimental conditions and their effect on the collision properties of the atoms. Experimentally, the standard method to create quasi-1D tubes is using a very strong laser field for the transversal direction compared to the lateral one \cite{kinoshita,haller}. This way the trap becomes highly anisotropic with the characteristic transversal length scale  $a_{\perp} \equiv \sqrt {\frac{\hbar}{M \omega_{\perp}}}$ much smaller than the longitudinal $a_{\parallel} \equiv \sqrt {\frac{\hbar}{M \omega_{\parallel}}}$ [$\omega_{\perp}$ ($\omega_{\parallel}$) is the transversal (longitudinal) harmonic confinement frequency]. Then the transverse degrees of freedom are energetically frozen to their ground states and the effective 1D interaction strength reads $g_{1D}= \frac{2\hbar^2 a_0}{M a^2_{\perp}} \left(1-\frac{|\zeta(1/2)| a_0}{\sqrt{2} a_{\perp}}\right)^{-1}$ \cite{olshanii}, where $a_0$ is the 3D s-wave scattering length. 

The 1D N-body Hamiltonian reads:
\begin{equation*}
H = - \frac{1}{2} \sum_{i=1}^N \frac{\partial^2}{\partial  x_i^2} + \sum_{i=1}^N \frac{x_i^2}{2} + g \sum_{i<j}\delta(x_i-x_j)
\end{equation*}
where the contact interaction of particles located at $x_i$, $i=1,...,N$ is represented by the Dirac $\delta$-function, and lengths and energies are scaled by $a_{\parallel}$ and $\hbar\omega_{\parallel}$ respectively. The single remaining parameter is thus the rescaled interaction strength  $g=g_{1D} \sqrt {M/\hbar^3\omega_{\parallel}}$ which is altered either by tuning $a_0$ via magnetic Feshbach resonances, or $a_{\perp}$ by modifying the transversal confinement. 

An important property of $H$ that we employ here is the separability $H=H_{CM}+H_r$, where $H_{CM} \equiv \frac{P^2}{2N}+\frac{N R^2}{2}$ describes a harmonic oscillation in the center-of-mass coordinates $R=\frac{1}{N} \displaystyle\sum_{i=1}^N x_i$ and $P=-i \displaystyle\sum_{i=1}^N \frac{\partial}{\partial x_i}$, while $H_r$ describes the relative motion of the particles. The relative motion is non-trivially coupled for any choice of relative coordinates for $N>2$, and will be exclusively addressed in the following  

\paragraph{Correlated pair wave function} 

The correlated pair wave function (CPWF) developed here, is inspired by the idea that the pairwise contact interaction may be adequately addressed in the many-body system, if the discontinuity that it causes is imposed on each pair of atoms in the ensemble, in a similar way as for a single pair. The CPWF for the relative motion is formed as a pairwise product expansion of functions based on the two-body solutions, thereby respecting the two exactly solvable limits of zero and infinite interaction strength for an arbitrary number of particles. Specifically, the construction principle is composed of the following three postulates:

(i) The CPWF for the relative motion of $N$ particles is a product of parabolic cylinder functions (PCF) $D_{\mu}$ \cite{abramowitz} of the distance $r_{ij}=|x_i-x_j|$ of each pair: 
\begin{equation}
\label{psi}
 \Psi_{\mathrm{cp}} = C \prod_{i<j}^{P}D_{\mu}(\beta r_{ij}),
\end{equation}
where $P = \frac{N(N-1)}{2}$ is the number of distinct pairs and the parameters $\beta$ and $\mu$, to be determined next, are identical for every pair since we deal with identical particles, while the absolute value enforces the bosonic permutation symmetry. The normalization constant $C$ will, without loss of generality, be omitted in the following.

(ii) We make the assumption
\begin{equation}
\label{beta}
 \beta=\sqrt{\frac{2}{N}},
\end{equation}
which ensures that $\Psi_{\mathrm{cp}}$ reproduces the exact analytically known solutions in the limits $g=0$ and $g \to \infty$, for any number of particles. In these limits, $\mu$ equals $0$ and $1$ respectively (this follows from the boundary condition imposed next), and in both cases (as well as for every integer $\mu$) $D_{\mu}(x)=e^{-\frac{x^2}{4}}\mathrm{He}_{\mu}(x)$ where $\mathrm{He}_{\mu}(x)$ are the modified Hermite polynomials. Therefore the total wave function $\Psi=\Psi_R  \Psi_r$ with $\Psi_R=e^{-\frac{R^2}{\beta^2}}$ and $\Psi_r=\Psi_{\mathrm{cp}}$ reproduces the noninteracting $\displaystyle\Psi= \exp \left(-\sum_{i=1}^{N} \frac{x_i^2}{2}\right)$ and infinitely repulsive limit $\displaystyle\Psi= \exp  \left(-\sum_{i=1}^{N} \frac{x_i^2}{2} \right)\prod_{i<j}^{P}|x_i-x_j|$ \cite{girardeau1}. The latter coincides with the (fermionic) determinant, written in form of pairs, and with implemented bosonic symmetry (Tonks-Girardeau limit). 

(iii) The boundary condition $2 \beta D'_{ij}(0)= g D_{ij}(0)$, where  $D'_{ij}=\frac{\partial D_{ij}}{\partial ( \beta r_{ij})}$ and $D_{ij} = D_{\mu}(\beta r_{ij})$, is imposed for each pair in the ensemble at $r_{ij}=0$, and determines $\mu$ for a certain value of the interaction strength $g$. With the known expressions for the PCF $D_{\mu}(x=0)=\frac{2^{\frac{\mu}{2}}\sqrt{\pi}}{\Gamma\left(\frac{1-\mu}{2}\right)}$ and $\frac{\partial D_{\mu}}{\partial x}(x=0)= -\frac{2^{\frac{\mu+1}{2}}\sqrt{\pi}}{\Gamma\left(-\frac{\mu}{2}\right)}$ \cite{abramowitz}, the resulting transcendental equation 
\begin{equation}
\label{bc}
\frac{g}{\beta}=-\frac{2^\frac{3}{2}\Gamma\left(\frac{1-\mu}{2}\right)}{\Gamma\left(\frac{-\mu}{2}\right)}
\end{equation}
is solved for $\mu$, selecting the solution in the interval $\mu \in [0,1]$ which corresponds to the ground state. The origin of this boundary condition will be discussed below. 

\begin{figure}
\includegraphics[width=4.0 cm,height=4.0 cm]{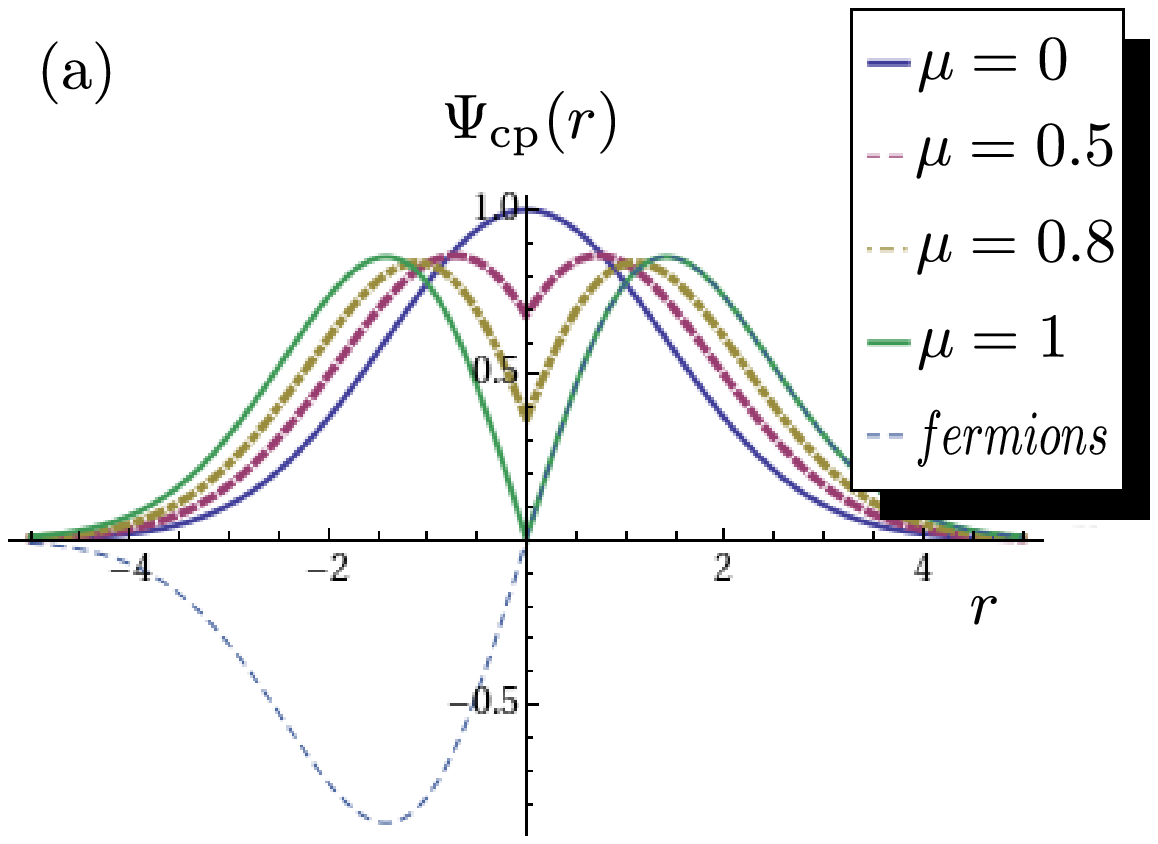}
\includegraphics[width=3.8 cm,height=3.8cm]{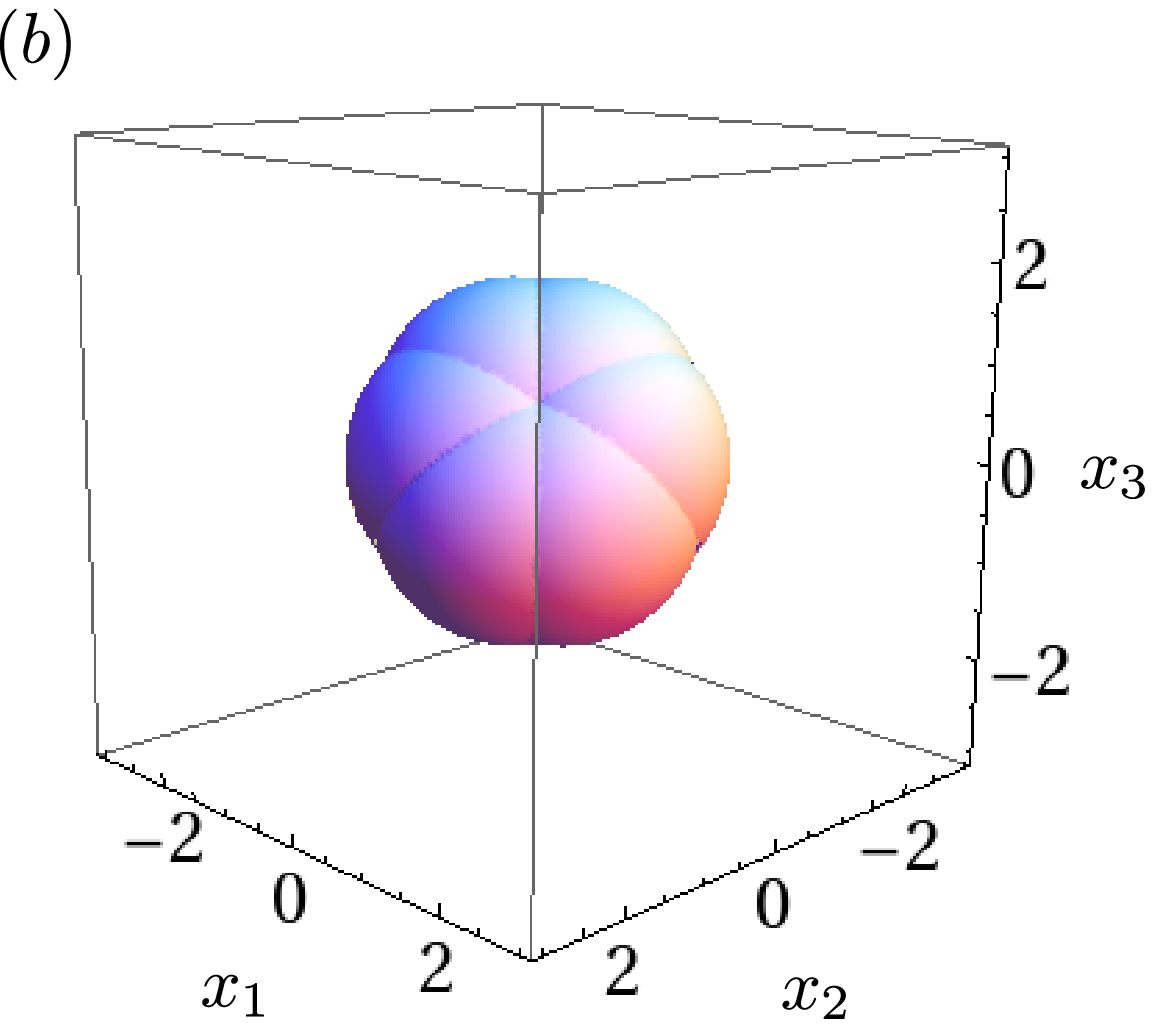}
\includegraphics[width=3.8 cm,height=3.8cm]{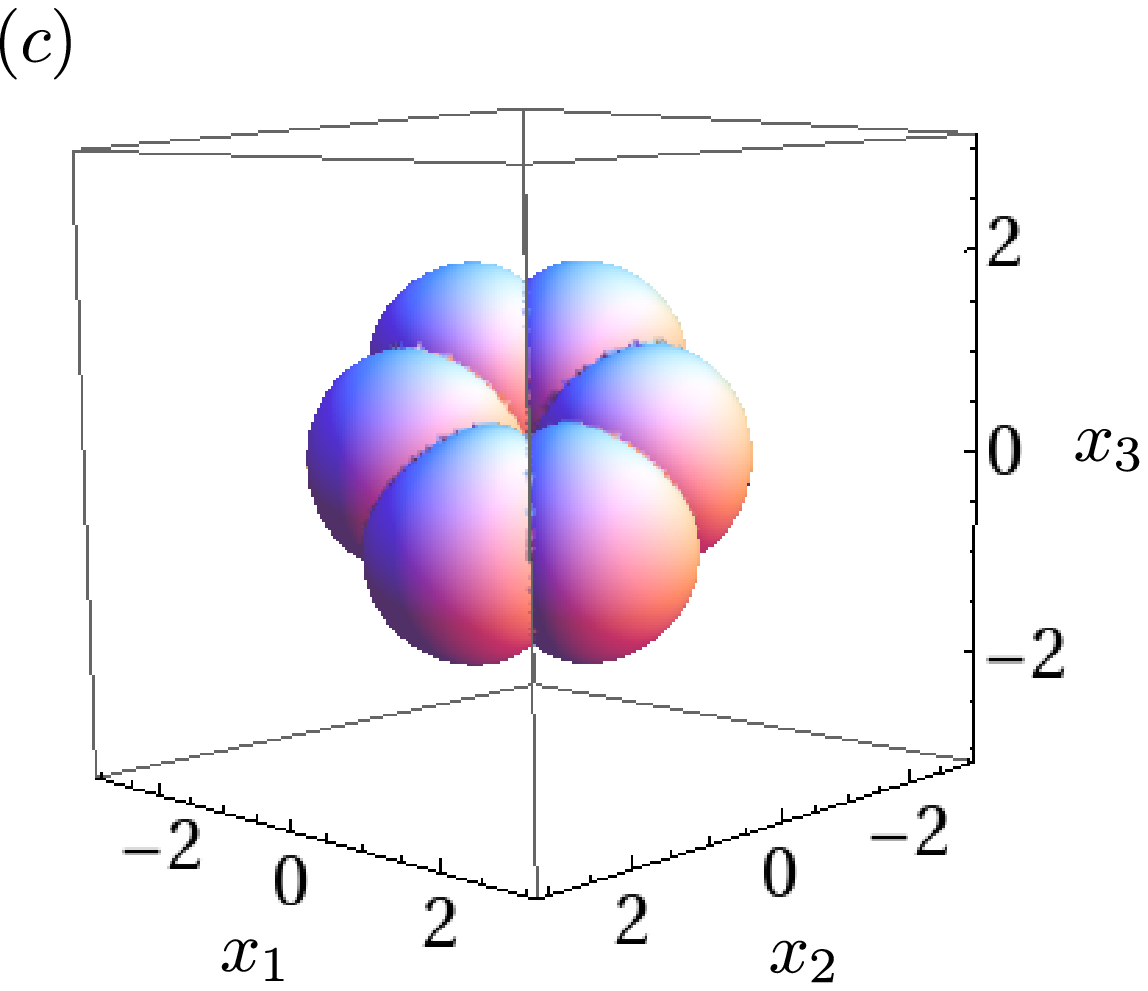}
\includegraphics[width=3.8 cm,height=3.8cm]{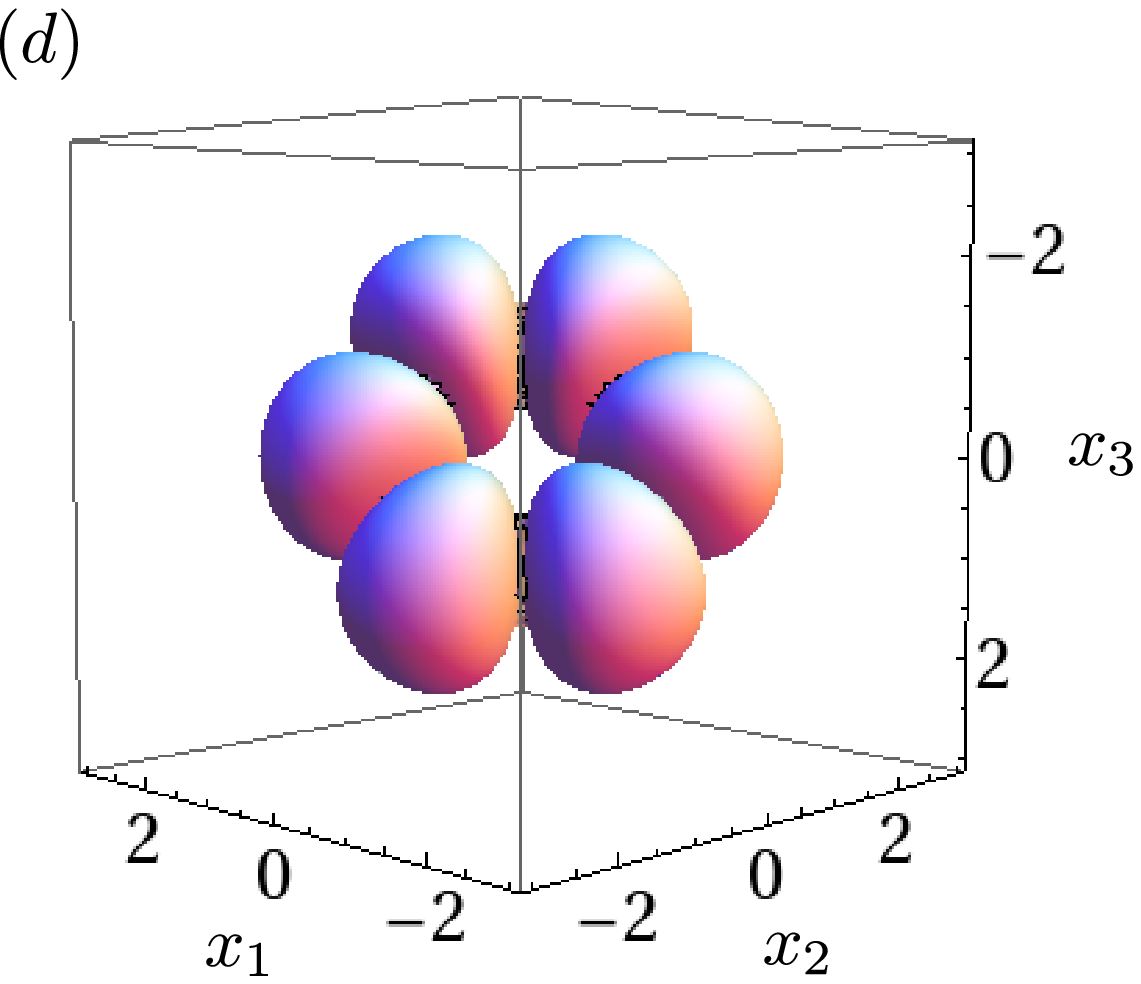}
\caption{(a) Wave function of the relative motion for two particles $\Psi_{\mathrm{cp}}(r \equiv x_1-x_2)$ and several strengths of the interaction ($\mu=0,0.5,0.8,1$) and comparison of $\mu=1$ with the fermionic antisymmetric function. Contour plots of the density distribution $|\Psi(x_1,x_2,x_3)|^2=0.01$  for three particles using the CPWF for the relative part for (b) $\mu=0.2$, (c) $\mu=0.5$, (d) $\mu=0.8$. }
\label{fig1}
\end{figure}

\paragraph{Discussion} 

Postulate (i) addressing the construction of a many-body wave function via a product of functions of the relative distance appears also in other treatments in similar form (see Ref. \onlinecite{book}). Nevertheless, the particular choice of the PCF as a building block proves to be, as we will show, a very efficient approach for the present problem, owing its inspiration to the two-body exact solution \cite{busch}. Key features of our approach are postulates (ii) and (iii) determining the properties of the PCF, thereby ensuring that $\Psi_{\mathrm{cp}}$  reproduces the two analytical limits of zero and infinite coupling for any $N$, and constitutes the exact solution at arbitrary interaction strength for $N=2$. Furthermore, in the two-particle case the analytical solution derived by Busch et. al \cite{busch} possesses a high degree of generality: it holds for bosons or fermions in one-, two- and three-dimensional harmonic traps and arbitrarily strong attractive or repulsive contact interactions, including ground and excited states of the relative motion. This readily implies that since our CPWF with this solution as a building block is sufficiently accurate for the 1D repulsive bosonic case, a similar Ansatz could be envisaged for other setups. 

To clarify the implied boundary condition in postulate (iii), as well as why the CPWF is exact for two particles we investigate the action of the relative motion Hamiltonian operator 
\begin{equation*} 
H_r= - \beta ^2 \sum_{i<j}^P\left(\frac{\partial}{\partial x_i}-\frac{\partial}{\partial x_j}\right)^2 + \beta^2 \sum_{i<j}^P\frac{r_{ij}^2}{4}+\sum_{i<j}^P g \delta(r_{ij}),
\end{equation*}
on $\Psi_{\mathrm{cp}}$. Introducing the notation $\displaystyle\phi^{kl,mn,...}= \prod_{i<j}D_{\mu}(\beta r_{ij \neq kl,mn,...})$ and $\chi_{lm}=1- 2 \delta_{lm} $, we obtain:
\begin{eqnarray} 
\label{hrpsi}
H_r\Psi_{\mathrm{cp}} &=& \sum_{i<j}^{P} \delta(r_{ij}) \left(-2 \beta D'_{ij} + g  D_{ij} \right)\phi^{ij}  \\
\nonumber
&-& \sum_{i<j}^{P} \left(D''_{ij} - \frac{(\beta r_{ij})^2}{4} D_{ij} \right)\phi^{ij}    \\
\nonumber
+ (1-\beta^2) \sum_{i<j}^{P} &D''_{ij} & \phi^{ij} + \beta^2 \sideset{}{'}\sum_{k,l,m,n} D'_{kl}D'_{mn} \chi_{lm} \phi^{kl,mn},
\end{eqnarray}
where $\displaystyle \sideset{}{'}\sum_{k,l,m,n}$ sums over non-repeated terms with $k \neq l \neq m \neq n$.
The first sum in Eq. (\ref{hrpsi}) which contains the $\delta$-interaction for each pair vanishes at $r_{ij}=0$ by imposing the boundary conditions $2 \beta D'_{ij}(0) = g  D_{ij}(0)$. This provides the behaviour at the two-body contact point arising from the discontinuity of $D'_{ij}$ for each pair. Therefore the manifolds $\mathcal{M}=\{(x_1,...x_i,...,x_j,...,x_N) \in \Re^N |x_i=x_j \}$ of contact (two particle collision) being of dimensionality $N-1$ are taken into account correctly, while higher order contact (three particle collision etc.) represent lower dimensional manifolds.  The second sum in Eq. (\ref{hrpsi}) is in the form of the Weber equation \cite{abramowitz}: $D''(x) - \frac{x^2}{4}D(x)=-(\mu +\frac{1}{2})D(x)$, for which the PCFs are solutions, and contributes to the energy by $P(\mu +\frac{1}{2})$. For $N=2$ ($\beta=1$, $P=1$) the last two terms in Eq. (\ref{hrpsi}) vanish and therefore $\Psi_{\mathrm{cp}}$ with the corresponding boundary condition is the exact solution with energy of the relative motion $\epsilon_r=(\mu +\frac{1}{2})$. For $N>2$ the last two sums in Eq. (\ref{hrpsi}) do not vanish and provide additional contributions to the energy. 

Before comparing our analytical approach with corresponding numerical results, we illustrate in Fig. \ref{fig1} the spatial distribution of two and three particles. For two particles the wave function of the relative motion $\Psi_{\mathrm{cp}}(r \equiv x_1-x_2)$ acquires a cusp at $r=0$ for $g>0$ [or $\mu>0$ from Eq. (\ref{bc})] which goes to zero as $g \to \infty$ ($\mu \to 1$), retrieving the fermionic state but with bosonic symmetry. Similarly, for three particles the contour plots demonstrate, for increasing interaction strength, depletion of the probability density along the collision manifolds ($r_{ij}=0$); the conceptually important physical insight offered by our CPWF approach is that it captures the correlation properties in the vicinity of collision surfaces, i.e., the tendency of the particles to repel and thus avoid each other.

\begin{figure}
\includegraphics[width=4.2 cm,height=4.2cm]{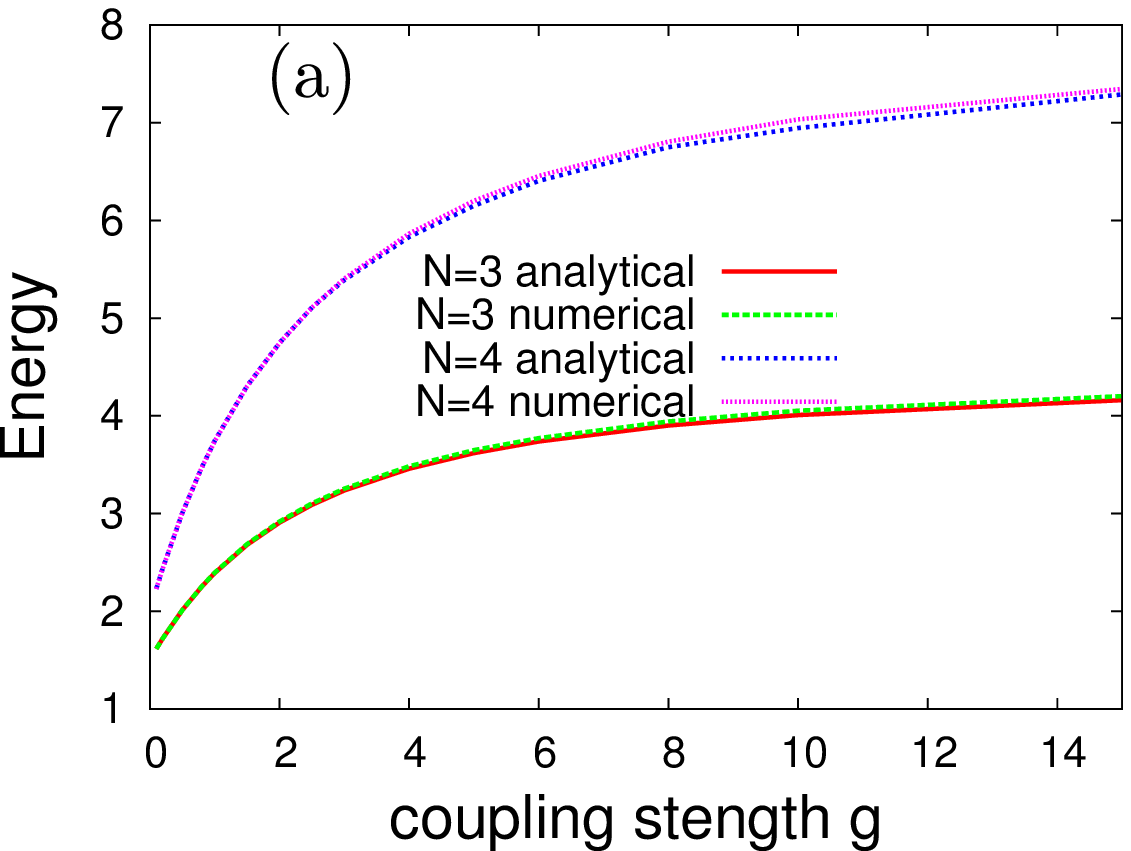}
\includegraphics[width=4.2 cm,height=4.2cm]{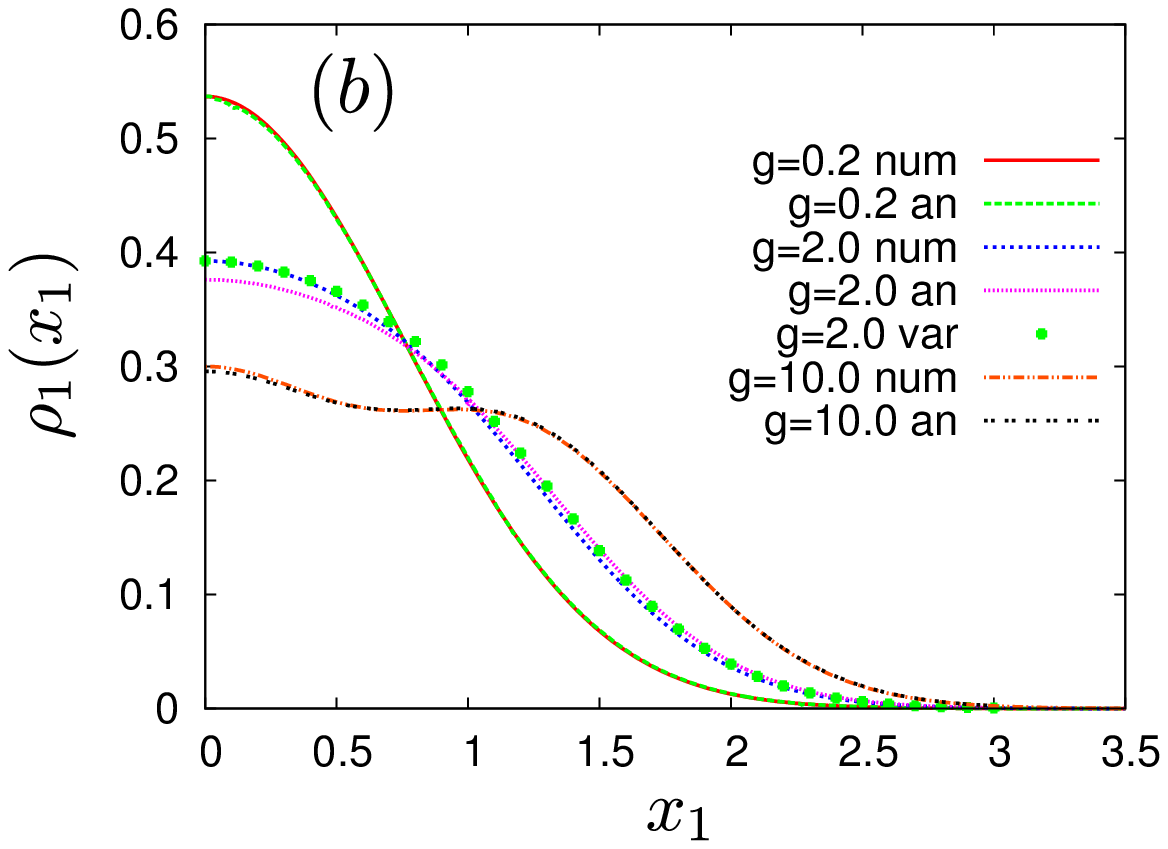}
\includegraphics[width=4.2 cm,height=4.2cm]{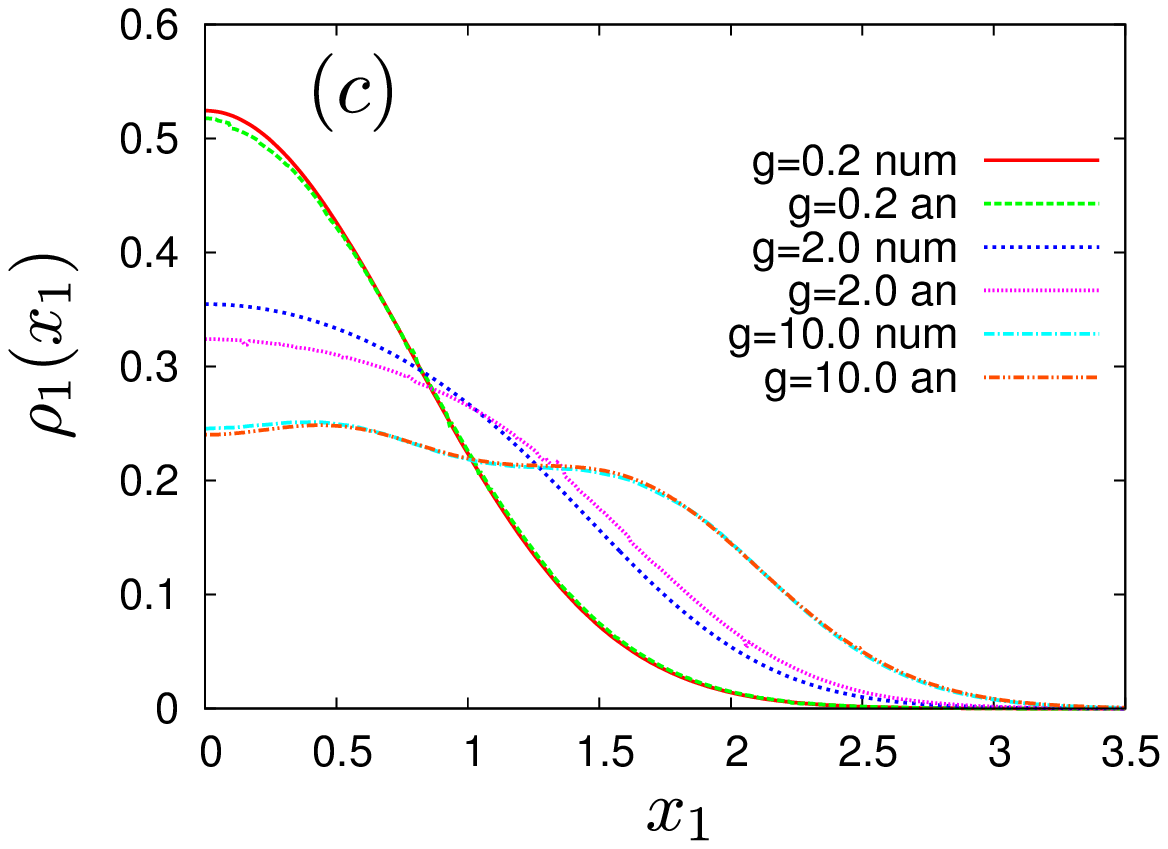}
\includegraphics[width=4.2 cm,height=4.2cm]{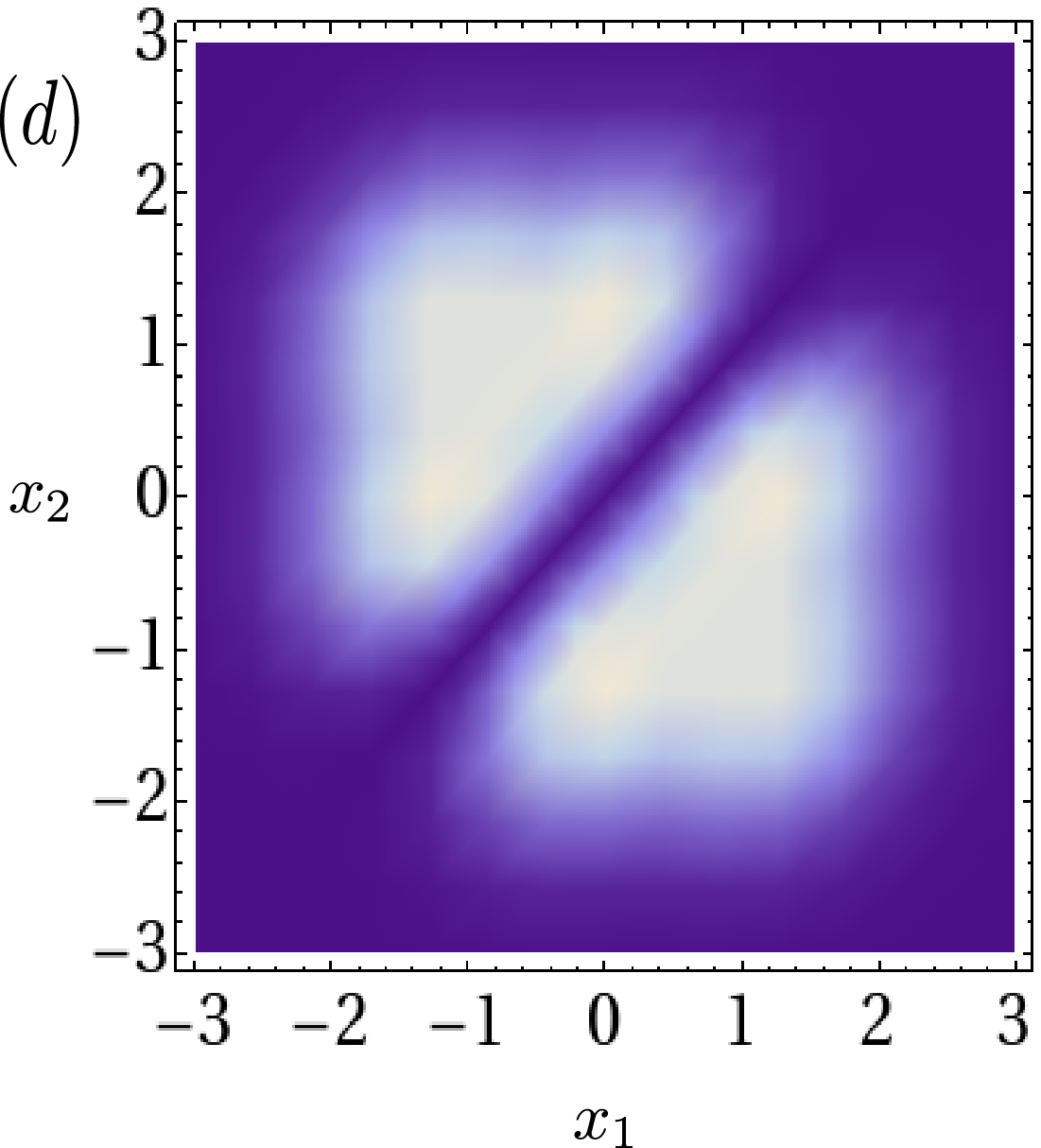}
\caption{(a) Energies as a function of the interaction strength $g$ obtained numerically and analytically via the CPWF. One-body densities $\rho_1(x_1)$, for (b) three and (c) four particles for interaction strengths $g=0.2,2.0,10.0$ comparing numerical and analytical results (only the $x_1>0$ part is shown since it is symmetric with respect to the ordinate). In (b) a comparison with a variational calculation for intermediate $g=2.0$ is shown. (d) Two-body density $\rho_2(x_1,x_2)$ for three particles analytically calculated for interaction strength $g=20.0$.}
\label{fig2}
\end{figure}

\paragraph{Accuracy of $\Psi_{\mathrm{cp}}$}

There are several numerical approaches to the problem which are limited in terms of the particle number they can handle for a given accuracy: Exact diagonalisation \cite{deuretzbacher}, Quantum Monte Carlo \cite{astrakharchik}, and multi-configurational approaches in terms of Hartree products \cite{sascha} have been employed. We will resort here to the latter method which is reliable for a small number of particles, assuring high numerical accuracy with a dense grid, small width of the Gaussian extrapolation of the $\delta$-function and large basis set (we use 500 basis functions and 30 single-particle orbitals). For details  of the computation of stationary states by improved relaxation (propagation in the imaginary time)  we refer the reader to Ref. \onlinecite{sascha} and references therein.  

The first observable we present in Fig. \ref{fig2} (a) for three and four particles is the expectation value of the energy with increasing interaction strength $g$. We observe excellent agreement between the energy calculated using CPWF and the numerical results, typically amounting to a relative accuracy of the order of $10^{-3}$ for intermediate values of $g$. Remarkably, for larger $g$ --a regime where convergence of numerical methods or approximative models is challenging \cite{dieter}-- the analytical calculation yields lower ground state energies, and thus proves to be more accurate. 

Next we analyze reduced density operators, namely one- and two- body densities, $\rho_1(x_1) \equiv \int_{-\infty}^{\infty} ... \int_{-\infty}^{\infty} |\Psi|^2  dx_2...dx_N$ and $\rho_2(x_1,x_2) \equiv \int_{-\infty}^{\infty} ... \int_{-\infty}^{\infty} |\Psi|^2  dx_3...dx_N $. The CPWF is shown to capture very well not only qualitatively, but also quantitatively, the properties of the crossover from weak to strong correlations. As a result of the increasing repulsion between the bosons, the one-body density [Fig. \ref{fig2} (b),(c)] flattens and forms an $N$-peak structure (including $x_1<0$) close to fermionization \cite {deuretzbacher, astrakharchik, sascha, girardeau1,dieter}. Note that for intermediate interaction ($g=2.0$) the deviation between CPWF and numerical results is somewhat larger; this is to be expected since $\Psi_{\mathrm{cp}}$ is exact in, and therefore very accurate close to, the limits $g=0$ and $g \to \infty$ ($\mu=0$ and $\mu \to 1$), while  between these limits ($\mu \approx 0.5$) the error of the analytical approach should be maximal. Even further improvement is obtained by relaxing postulate (ii) and performing  a variational optimization of $\beta$ instead of keeping it fixed at $\sqrt{\frac{2}{N}}$ [Eq. (\ref{beta})]. For optimal $\beta_{\mathrm{var}} \approx 0.84$ the energy is slightly lower by $0.1 \%$ and the one-body density fits excellently to the numerical one [see Fig. \ref{fig2} (b)]. Properties of the two-body density  such its depletion along the diagonal $x_1=x_2$, due to the avoided simultaneous contact (a behavior attributed to the cusps of the CPWF in Fig. \ref{fig1}), and peaks on the off-diagonal  are shown in Fig. \ref{fig2} (d) for three strongly interacting particles in agreement with the results in Refs. \onlinecite {sascha,girardeau1}. 

\paragraph{Conclusions}

The many-body wave function presented in this letter for the problem of zero-range repulsively interacting bosons with arbitrary coupling strength in a one-dimensional harmonic trap, has been inspired by exact analytical solution of the two-body case. We have shown that taking a product expansion of pair wave functions similar to the two-body solution, but modified such that the two analytically known limits of zero and infinite interaction strength are exactly reproduced for any number of particles, leads to an impressive agreement with extensive numerical calculations for any value of the coupling strength. Therefore the construction principle of our approach may be applicable to other setups, e.g. in higher dimensions, where the building block -the two-particle function- can be represented by an analytical expression similar to the exact solution of the relative motion for the two-body problem, and some limiting cases of the relevant parameters (here e.g. zero and infinite interaction strength) are either exactly or approximately known. Improvements of our approach may be possible (e.g. by taking into account lower dimensional spaces where three or more particles meet), and it is an excellent starting point for variational calculations (performed also in an exemplary case here) or for Quantum Monte-Carlo methods as an appropriate guiding function of Jastrow type, or for exploring the limits of mean-field treatments. Along with an accurate description of a highly correlated many-body problem, we believe that our approach offers valuable physical insight into the correlation properties at collision manifolds, which is conceptually useful in the theoretical treatment of many-body physics. 

\paragraph{Acknowledgements}
The authors are thankful to F.K. Diakonos and C. Morfonios for valuable discussions. Financial support by the Deutsche Forschungsgemeinschaft is acknowledged.


\begin{thebibliography}{10}

\bibitem{pethick_smith}
C.~J. Pethick and H. Smith, {\em Bose-Einstein condensation in dilute gases} (Cambridge University Press, Cambridge, 2008); L. Pitaevskii and S. Stringari, {\em Bose-Einstein Condensation} (Oxford   University Press, Oxford, 2003).

\bibitem{bloch}
I. Bloch, J. Dalibard, and W. Zwerger, Rev. Mod. Phys. {\bf 80},  885  (2008). 

\bibitem{chin}
C. Chin et al., Rev. Mod. Phys. {\bf 82}, 1225 (2010). 

\bibitem{greiner}
M. Greiner et al., Nature {\bf  415},  39  (2002).

\bibitem{olshanii}
M. Olshanii, Phys. Rev. Lett. {\bf 81},  938  (1998).

\bibitem{Pitaevskii}
L. P. Pitaevskii, Zh. Eksp. Teor. Fiz. {\bf 40} 646 (1961) [Sov. Phys. JETP  {\bf 13} 451 (1961)].

\bibitem{lieb}
E.~H. Lieb and W. Liniger, Phys. Rev. {\bf 130},  1605  (1963).

\bibitem{hao}
Y. Hao et al., Phys. Rev. A {\bf 73},  063617 (2006).

\bibitem{girardeau}
M. Girardeau, J. Math. Phys. {\bf 1},  516  (1960).

\bibitem{kinoshita}
T. Kinoshita, T. Wenger, and D.~S. Weiss, Science {\bf 305},  1125  (2004); B. Paredes et al., Nature {\bf 429},  277  (2004).

\bibitem{haller}
E. Haller et al., Science {\bf 325}, 1224 (2009).

\bibitem{abramowitz} 
M. Abramowitz and I. A. Stegun  {\em Handbook of Mathematical Functions} (Dover, New York, 1972).

\bibitem{girardeau1}
M.~D. Girardeau, E.~M. Wright, and J.~M. Triscari, Phys. Rev. A {\bf 63}, 033601  (2001).

\bibitem{book}
J. Rychlewski, {\em Explicitly Correlated Wave Functions in Chemistry and Physics: Theory and Applications} (Springer Netherlands, 2003)

\bibitem{busch}
T. Busch et al., Found. Phys. {\bf 28}, 549  (1998).

\bibitem{deuretzbacher}
F. Deuretzbacher et al., Phys. Rev. A {\bf  75},  013614  (2007).

 \bibitem{astrakharchik}
G. E. Astrakharchik et al.,  J. Phys. B {\bf 37 } 205 (2004).

\bibitem{sascha}
S. Z{\"o}llner, H.-D. Meyer, and P. Schmelcher, Phys. Rev. A {\bf 74},  063611, 053612 (2006).

\bibitem{dieter}
 T. Ernst et al.,   arXiv:1104.2627v1 


\end{thebibliography}
\end{document}